\documentclass[12pt]{article}

\expandafter\let\csname equation*\endcsname\relax
\expandafter\let\csname endequation*\endcsname\relax
\usepackage{amsmath}

\usepackage[T1]{fontenc}
\usepackage{textcomp}

\usepackage{amsfonts}

\usepackage{graphicx} 
\usepackage[active]{srcltx}
\title{Electromagnetic Waves in a Uniform Gravitational Field and Planck's Postulate}
\author{Luis Acedo\thanks{Corresponding author e-mail: luiacrod@imm.upv.es} and Michael M.\ Tung\\
Instituto de Matem\'atica Multidisciplinar, \\
Universitat Polit\`ecnica de Val\`encia, \\ Camino de Vera s/n, 46022 Valencia, Spain}

\begin{document}
\maketitle

\begin{abstract}
The gravitational redshift forms the central part of the majority of the classical tests
for the general theory of relativity. It could be successfully checked even in laboratory
experiments on the earth's surface. The standard derivation of this effect is based on the
distortion of the local structure of spacetime induced by large masses. The resulting
gravitational time-dilation near these masses gives rise to a frequency change of any
periodic process, including electromagnetic oscillations as the wave propagates across
the gravitational field. This phenomenon can be tackled with classical electrodynamics
assuming a curved spacetime background and Maxwell's equations in a generally covariant form.
In the present paper, we show that in a classical field-theoretical context the gravitational
redshift can be interpreted as the propagation of electromagnetic waves in a medium with corresponding
conductivity $\sigma=g/(\mu_0 c^3)$, where $g$ is the gravitational acceleration and $\mu_0$
is the vacuum magnetic permeability. Moreover, the energy density of the wave remains
proportional to its frequency in agreement with Planck's postulate.
\end{abstract}



\section{Introduction}
\label{Sect_I}

The first classical test of the general theory of relativity in a terrestrial environment 
was devised and carried out by Robert Pound and Glen A.\ Rebka, Jr., in 1959~\cite{poundrebka}.
The idea of this experiment is quite simple: An atom of iron-$57$ emits by radioactive decay
a gamma photon with an energy of $14$~keV. Such a photon travels upward (in this case from the
basement of Jefferson Laboratory at Harvard's University) until it reaches a certain height $h$
(in the original experiment at the attic of the building it was $h=22.5$~m) where its energy
has changed to
\begin{equation}
E_1=E_0 \left(1-\displaystyle\frac{g h}{c^2}\right),
\end{equation}
where $g$ denotes the strength of the gravitational field, being equivalent to the
acceleration of objects under its influence. The interpretation of the redshift in
a static gravitational field is not as simple as it appears; for a detailed discussion
see Ref.~\cite{okun}.

At the maximum height of its trajectory,
the redshifted photon can no more be absorbed by the iron-$57$ atoms in the receiver.
In a moment of inspiration, Pound and Rebka realised that the gravitational redshift
of the photon can be cancelled out by an artificially created Doppler blueshift equivalent
to a downward motion towards the basement:
\begin{equation}
E_2 = E_1 \left(1+\displaystyle\frac{v}{c}\right)
    = E_0 \left(1+\displaystyle\frac{v}{c}
          -\displaystyle\frac{g h}{c^2}+{\cal O}\left(c^{-3}\right)\right).
\end{equation}
As can be clearly seen, the cancellation with $E_2 \approx E_0$ takes place if we choose
the velocity $v=g h/c$. However, the fractional energy change
$(E_1-E_0)/E_0 =2.5 \times 10^{-15}$
is so small that the experiment would be contaminated by atom recoiling. Fortunately, just
a year before, Rudolf L.\ M\"ossbauer had discovered that atoms belonging to a solid lattice
share the recoiling momentum and, consequently, the associated energy and velocity change is
negligible~\cite{frauenfelder}. By using M\"ossbauer spectroscopy, Pound and Rebka were then
able to measure the velocity necessary to counteract the gravitational redshift by the Doppler
effect, $v=g h/c \approx  7.5 \times 10^{-4}$~mm/s. These very small velocities 
were obtained by placing the sample on a conical speaker membrane reproducing a
low-frequency sound between $10$ and $50$~Hz. Note that as a sequel to this initial
experiment, many higher-precision tests were added, {\it e.g.}\ in 1980 by means of a
space-borne hydrogen maser~\cite{vessot}.

From a historical point of view, already in 1784, John Michell, an English philosopher and
geologist, anticipated the gravitational weakening of starlight in a letter to Henry
Cavendish~\cite{schaffer}. Michell's calculations were based on the
Newtonian corpuscular theory of light which was later on rejected after the arrival of the
wave theory in the early XIXth century.

The objective of this paper is to analyse the propagation of light in a gravitational
field by means of the covariant Maxwell equations with a curved spacetime background.
We will consider a uniform
gravitational field and a uniform accelerating frame as proposed by Desloge~\cite{deslogefield}
and assume the following line element:
\begin{equation}\label{lineelement}
ds^2 = -\alpha^2(z)\,c^2 dt^2 + dx^2 + dy^2 + dz^2,
\end{equation}
where time-dilation is included either by considering a uniform gravitational field or
by a uniform accelerating frame along the $z$ axis:
\begin{equation}
\label{alpha}
\alpha(z) = \left\{
\begin{array}{rcl}
    e^{g z /c^2} & \quad & \mbox{uniform gravitational field (UGF)} \\
    \noalign{\smallskip}
    1+\displaystyle\frac{g z}{c^2} & \quad & \mbox{uniform accelerating rigid frame (UAF)}.
\end{array}
\right.
\end{equation}
The UGF metric is a straightforward solution of the condition that the
initial local acceleration of a particle must have the same value at all
points in order to describe a uniform field. The corresponding differential equation
therefore is:
$$
\frac{1}{\alpha}\frac{d\alpha}{dz} = \frac{g}{c^2},\quad\mbox{with}\quad\alpha(0)=1.
$$
On the other hand, the UAF metric for a uniformly accelerating rigid frame in field-free
space assumes an underlying flat space, which yields the following simple
differential equation and boundary conditions:
$$
\frac{d^2\alpha}{dz^2} = 0,\qquad
\frac{d\alpha}{dz}(0) = \frac{g}{c^2},\qquad
\alpha(0) = 1.
$$
It is not difficult to see that the particular expressions for $\alpha(z)$ in Eq.~\eqref{alpha}
are just the exact solutions of these two differential systems, corresponding to the UGF and
UAF case, respectively.

The UAF metric describes an underlying flat spacetime, whereas the UGF metric represents
curved spacetime as expected. So in principle, the two approaches of Eq.~\eqref{alpha} are
fundamentally different and were used by Desloge to explicitly show that the observations
made in a UGF are not strictly identical to those of a UAF.
Here, the statement of the principle of equivalence between acceleration and gravity is only
valid as a heuristic approximation. In this context, the two different metrics also allowed
to analyse and discuss gravitational redshift effects in a straightforward manner.

In our opinion, it would be pedagogical and helpful to gain new insights if this approach
were also extended to include the effects of gravitational redshift from the point of view of
classical electrodynamics in combination with general relativity. Classical references on the
subject arrive at the covariant formulation of Maxwell's equations and stop there. Only in some
of the more advanced textbooks particular solutions are fully discussed. We will show that
Desloge's metric is an excellent testing ground by obtaining explicit solutions of 
Maxwell equations for electromagnetic waves in a curved spacetime background.
Furthermore, we will see how the conceptual transition from classical electrodynamics to its
extension in general relativity is minimised in this approach. 

The paper is organised as follows: In Section~\ref{Sect_II} we first set up the
Maxwell equations for the UGF and UAF metrics. In particular, we study the
electrostatic field of an infinite and uniformly charged plate in a uniform gravitational
field and derive its solution. Next, in Section~\ref{Sect_III}, the electromagnetic wave
equation is formulated for a general UGF frame. An approximate analytical solution of the
wave equation is also calculated and expressed in terms of the two linearly independent
solutions $\mathop{\rm Ai}(z)$ and $\mathop{\rm Bi}(z)$
of the Airy equation $y''(x)-xy=0$ extended to the complex plane by
analytic continuation. It can be shown that the relation among the
energy density of the wave and its frequency satisfies Planck's postulate.
Section~\ref{Sect_IV} concludes the paper with some final remarks and observations.

\section{Covariant Maxwell Equations in a Uniform Gravitational Field}
\label{Sect_II}

In $4$-dimensional spacetime, the Faraday tensor, or covariant electromagnetic field tensor,
allows the physical laws which govern electromagnetic phenomena to be written in a very concise
form. For an underlying metric with signature $(-,+,+,+)$, it is defined by 
\begin{equation}
F_{\mu \nu}=\left( \begin{array}{cccc} 
0 & -E_x/c & -E_y/c & -E_z/c \\
\noalign{\smallskip}
E_x/c & 0 & B_z & -B_y \\
\noalign{\smallskip}
E_y/c & -B_z & 0 & B_x \\
\noalign{\smallskip}
E_z/c & B_y & -B_x & 0
\end{array}\right),
\end{equation}
where as usual the electromagnetic field is decomposed into the
field vectors $\mathbf{E}=(E_x,E_y,E_z)$ and $\mathbf{B}=(B_x,B_y,B_z)$ as seen in a
frame of a particular observer. Using Desloge's approach, the inverse of $F_{\mu\nu}$
with the metric of Eq.~\eqref{lineelement} is given in contravariant form by
\begin{equation}
\label{Fcont}
F^{\mu \nu}=\left( \begin{array}{cccc} 
0 & E_x/c\alpha^2(z) & E_y/c\alpha^2(z) & E_z/c\alpha^2(z) \\
\noalign{\smallskip}
-E_x/c\alpha^2(z) & 0 & B_z & -B_y \\
\noalign{\smallskip}
-E_y/c\alpha^2(z) & -B_z & 0 & B_x \\
\noalign{\smallskip}
-E_z/c\alpha^2(z) & B_y & -B_x & 0
\end{array}\right).
\end{equation}
Maxwell's equations can then be recast in covariant form. For this purpose the source equations
of the electric and magnetic fields are summarised in a single relation containing the covariant
derivative of the electromagnetic field tensor:
\begin{equation}
\label{sourceeq}
F^{\mu\nu}{}_{;\nu}=\mu_0 j^\mu,
\end{equation}
where the semicolon denotes the covariant derivative and $j^\mu=(c \rho, j_x, j_y, j_z)$
is the current four-vector. The rotational equation for the electric field and the divergence-free
condition of the magnetic field are incorporated in the cyclic equation for $F_{\mu\nu}$:
\begin{equation}
\label{cycliceq}
F_{[\lambda \mu ; \nu]} = 0
\quad\mbox{or}\quad
F_{\lambda \mu ; \nu}+F_{\mu \nu; \lambda}+F_{\nu \lambda; \mu} = 0,
\end{equation}
where the cyclic permutations of the indices may be abbreviated by the common bracket notation. 
This implies that in all terms the covariant derivatives, which include Christoffel symbols,
cancel out, and we can replace the covariant derivatives by ordinary derivatives obtaining
\begin{equation}
\label{cyclic}
F_{[\lambda \mu , \nu]} = 0
\quad\mbox{or}\quad
F_{\lambda \mu , \nu}+F_{\mu \nu, \lambda}+F_{\nu \lambda, \mu} = 0,
\end{equation}
where the comma denotes now conventional partial derivatives. After expanding the covariant
derivative in Eq.~(\ref{sourceeq}), we also find
\begin{equation}
\label{sourceexp}
F^{\mu \nu}{}_{, \nu} +
\displaystyle\frac{\left(\sqrt{-g}\right)_{, \nu}}{\sqrt{-g}}
F^{\mu \nu}=\mu_0 j^\mu,
\end{equation}
where we have taken into account that the contraction of the symmetric Christoffel symbols with
the antisymmetric electromagnetic tensor is zero in the absence of torsion, namely
$\Gamma_{\alpha \beta}^\mu F^{\alpha \beta}=0$. In Desloge's approach, the metric tensor is
given by
\begin{equation}
\label{metric}
g_{\mu \nu}=\left(\begin{array}{cccc}
-\alpha^2(z) & 0 & 0 & 0 \\
\noalign{\smallskip}
0 & 1 & 0 & 0 \\
\noalign{\smallskip}
0 & 0 & 1 & 0 \\
\noalign{\smallskip}
0 & 0 & 0 & 1 
\end{array}\right),
\end{equation}
and consequently, $\det\left(g_{\mu\nu}\right)=-\alpha^2(z)$.
From Eqs.~\eqref{Fcont} and (\ref{sourceexp}) we arrive after some simplification at the
following conditions for the electric field
\begin{equation}
\label{divE}
\mathbf{\nabla} \cdot \mathbf{E} - \displaystyle\frac{\dot{\alpha}}{\alpha} E_z
= \displaystyle\frac{\rho}{\varepsilon_0}\,\alpha^2(z),
\end{equation}
and similarly for the magnetic field
\begin{equation}
\label{nablaB}
\mathbf{\nabla} \times \mathbf{B} =
\mu_0\,\mathbf{j} + \displaystyle\frac{\dot{\alpha}}{\alpha}\,\mathbf{B} \times \hat{\mathbf k}
+ \displaystyle\frac{1}{c^2 \alpha^2} \displaystyle\frac{\partial \mathbf{E}}{\partial t},
\end{equation}
where $\hat{\mathbf k}$, as usual, denotes the unit vector in $z$-direction.
In an analogous way, from Eq.~(\ref{cycliceq}) we may derive the two remaining Maxwell equations:
\begin{eqnarray}
\label{nablaE}
\mathbf{\nabla} \times \mathbf{E} &=& -\displaystyle\frac{\partial \mathbf{B}}{\partial t}, \\
\noalign{\smallskip}
\mathbf{\nabla} \cdot \mathbf{B} &=& 0.
\end{eqnarray} 
Before starting to find the full electromagnetic wave solution of this system, it is useful
to study the simpler electrostatic case. For this purpose, consider in a particular UGF frame
an infinite metallic plate which is uniformly charged and is perpendicular to the $z$-axis.
From Eqs.~(\ref{divE}) and (\ref{nablaE}) we find that the electric field is irrotational and
satisfies
\begin{equation}
\label{nablaEplate}
\mathbf{\nabla} \cdot \mathbf{E}-\displaystyle\frac{g}{c^2}E_z=0,
 \quad (\mbox{outside the charged plate})
\end{equation}
which indicates that the translational Poincar\'e symmetry has been broken by gravitation.
As the electric field is aligned in the $z$-direction, in the vacuum Eqs.~(\ref{divE}) and
(\ref{nablaEplate}) reduce to
\begin{eqnarray}
\displaystyle\frac{\partial E_z}{\partial z}-\displaystyle\frac{g}{c^2} E_z &=&0, \quad \mbox{(UGF)} \\
\noalign{\smallskip}
\displaystyle\frac{\partial E_z}{\partial z}-\displaystyle\frac{g/c^2}{1+g z/c^2}E_z &=&0, \quad \mbox{(UAF)}  
\end{eqnarray}
which can be readily integrated to yield
\begin{eqnarray}
E_z(z) &=& E_0\,e^{z/L}, \quad \mbox{(UGF)}
\label{EUGF} \\
\noalign{\smallskip}
E_z(z)&=& E_0 \left(1+ \displaystyle\frac{z}{L} \right).\quad \mbox{(UAF)}
\label{EUAF}
\end{eqnarray}
Here $L=c^2/g$ is a characteristic length scale associated with the gravitational field.
For weak gravitational fields this scale is very large: Assuming, for example,  a uniform
gravitational field with an acceleration corresponding to the local acceleration at the
surface of the earth, namely $g= 9.8\mbox{ m}/\mbox{s}^2$, produces the scale value
$L \approx 0.97\,\mbox{lyr}$. The length scale in general also provides for an estimate in which
domain the UGF and UAF descriptions agree or differ in their predictions. In the domain $z \ll L$
expression Eq.~\eqref{EUGF} converges to Eq.~\eqref{EUAF}, so that both results agree 
to first order:
\begin{equation}
\label{Ezapprox}
E_z(z)=E_0\left[ 1+\displaystyle\frac{z}{L}
 + {\cal O}\left(\left(\displaystyle\frac{z}{L}\right)^2\right)\right].
\end{equation}
As it has been shown for other phenomena, a
uniformly accelerated rigid frame and a uniform gravitational field are not strictly equivalent
on larger scales, although the equivalence principle, which guided Einstein heuristically
towards a formulation of general relativity, is still valid locally for weak fields and small
accelerations~\cite{deslogefield,deslogeredshift}.

It is worthwhile to note that the gravitational acceleration corresponding to
Eqs.~\eqref{EUGF} and \eqref{EUAF} are not exactly realised in nature. In fact, the discrepancy
between the two alternatives (UGF or UAF) occurs outside the physical domain. The physical domain
is fixed by the scale $L=c^2/g$ such that $|z|\ll L$, since the spacetime metric of the UAF
description strictly applies only in the limit $gz/c^2\ll 1$,
and furthermore the UGF approach becomes problematic in the sufficiently large-$z$ domain.
In any case, the full analytical treatment of the considerably more complicated case with a
black-hole background spacetime, which represents a physical and very strong gravitational field
in its vicinity, deserves special attention and is planned in a future work. Nevertheless, the
current approach serves as a viable and instructive guide to explore wave-like solutions of
Maxwell's solutions and their energy content for uniformly accelerating frames.

\section{Electromagnetic waves in a uniform gravitational field}
\label{Sect_III}
Considering Desloge's UGF metric of Eq.~\eqref{alpha}, which implies a uniform gravitational
acceleration along the $z$-axis, Maxwell equations in vacuum take the following form:
\begin{subequations}
\begin{eqnarray}
\mathbf{\nabla} \cdot \mathbf{E}&=&\displaystyle\frac{g}{c^2}E_z,  
\label{divE2}
\\
\noalign{\smallskip}
\mathbf{\nabla} \times \mathbf{E}&=& -\displaystyle\frac{\partial \mathbf{B}}{\partial t},
\label{curlE} \\
\noalign{\smallskip}
\mathbf{\nabla} \cdot \mathbf{B} &=& 0, \\
\noalign{\smallskip}
\mathbf{\nabla} \times \mathbf{B} &=&
\displaystyle\frac{g}{c^2} \mathbf{B} \times \hat{\mathbf k}
+\displaystyle\frac{1}{c^2} e^{-2 g z / c^2}\displaystyle\frac{\partial \mathbf{E}}{\partial t}.
\label{curlB}
\end{eqnarray}
\end{subequations}
In order to derive the wave equation, we apply the standard technique in classical electrodynamics
by taking the curl of Eq.~\eqref{curlE} and thereby obtain
\begin{equation}
\label{rotrotE}
\mathbf{\nabla}\times(\mathbf{\nabla} \times \mathbf{E})=\mathbf{\nabla}\left(\mathbf{\nabla}\cdot \mathbf{E}\right)-\nabla^2 \mathbf{E}=-\displaystyle\frac{\partial}{\partial t} 
\mathbf{\nabla} \times \mathbf{B}.
\end{equation}
By direct substitution of the expressions for the divergence of the electric field,
Eq.~\eqref{divE2}, and the curl of the magnetic field, Eq.~\eqref{curlB}, we finally get
\begin{equation}
\label{Ewave}
\nabla^2 {\bf E} -
\displaystyle\frac{g}{c^2}\displaystyle\frac{\partial}{\partial t}\mathbf{B}\times\hat{\mathbf k}
-\displaystyle\frac{1}{c^2} e^{-2 g z/c^2} \displaystyle\frac{\partial^2 \mathbf{E}}{\partial t^2}
-\displaystyle\frac{g}{c^2} \mathbf{\nabla} E_z = 0.
\end{equation}
Proceeding in a similar way, we take the curl of the curl of the magnetic field and simplify
by using the remaining Maxwell equations to arrive at
\begin{equation}
\label{rotrotB}
\nabla^2 \mathbf{B}+\displaystyle\frac{g}{c^2}\displaystyle\frac{\partial\mathbf{B}}{\partial z}
+\displaystyle\frac{2 g}{c^4} \displaystyle\frac{\partial}{\partial t}
\mathbf{E} \times \hat{\mathbf k}
-\displaystyle\frac{1}{c^2} e^{-2 g z/c^2}
  \displaystyle\frac{\partial^2 \mathbf{B}}{\partial t^2}=0.
\end{equation}
Equations (\ref{rotrotE}) and (\ref{rotrotB}) are apparently quite different from the well-known
wave equations. In order to explain the behaviour of electromagnetic waves in the UGF system,
we make some simplifying assumptions:
\begin{enumerate}
\item[(i)] The wave travels upwards or downwards, parallel to the $z$-axis. 
\item[(ii)] The are no longitudinal electromagnetic components: $E_z=0$, $B_z=0$.
\item[(iii)] The electric, magnetic and propagation vector satisfy the standard
	     right-hand-rule which implies
	     $\mathbf{E}=c \mathbf{B}\times\hat{\mathbf k}$
	     and $\mathbf{B}=-\mathbf{E}\times\hat{\mathbf k}/c$.
\end{enumerate}
With these conditions, Eq. (\ref{Ewave}) becomes
\begin{equation}
\label{Etelegraph}
\nabla^2 \mathbf{E}-\displaystyle\frac{g}{c^3} \displaystyle\frac{\partial \mathbf{E}}{\partial t}-\displaystyle\frac{1}{c^2} e^{-2 g z/c^2} \displaystyle\frac{\partial^2 
\mathbf{E}}{\partial t^2}=0.
\end{equation}
Except for the additional factor of the second-order time derivative, this equation coincides
with the telegraph equation for the propagation of electromagnetic waves in a conducting
medium~\cite{panofsky}. As in the telegraph equation, we identify in Eq.~\eqref{Etelegraph}
the coefficient of the first time derivative with $\mu_0\sigma$, so that the conductivity of
the gravitational field can be taken as $\sigma= g/(\mu_0 c^3)$. We now propose the following
general solution for the complex electric field:
\begin{equation}
\label{proposal}
\mathbf{E}=\mathbf{E}_0 e^{\gamma(z)-i \omega t},
\end{equation}
where $i$ is the imaginary unit, $\omega$ is the frequency of the wave, $t$ is 
the coordinate time, and $\gamma(z)$ is a
function still to be determined. Observe that, if we use coordinate time instead of
local time at a fixed spatial position, the frequency measured is constant. By inserting
Eq.~(\ref{proposal}) into Eq.~(\ref{Etelegraph}) we obtain
\begin{equation}
\label{gammaz}
\displaystyle\frac{d^2 \gamma}{d z^2}+\left( \displaystyle\frac{d \gamma}{d z} \right)^2
+ \displaystyle\frac{\omega^2}{c^2} e^{-2 q z} + i \displaystyle\frac{q \omega}{c}=0,
\end{equation}
where $q=g/c^2$ is the inverse of the characteristic length of the gravitational field.
Equation (\ref{Etelegraph}) is a second-order non-linear differential equation which,
fortunately, can be linearised by the variable change
$\gamma(z)=\ln\left(z+\mathcal{A}(z)\right)$, with the new unknown function $\mathcal{A}(z)$.
It then follows that
\begin{equation}
\label{Aeq}
\displaystyle\frac{d^2 \mathcal{A}}{d z^2}
+\left(\displaystyle\frac{\omega^2}{c^2} e^{-2 q z}
+ i \displaystyle\frac{q \omega}{c}\right) \left(\mathcal{A}+z\right)=0.
\end{equation}
This differential equation for $\mathcal{A}(z)$ is non-homogeneous, however one of its
particular solutions is simply $\mathcal{A}_p(z)=-z$. The general solution of the homogeneous
equation, $\mathcal{A}_h(z)$, can be expressed in terms of Bessel functions with complex index
and the gamma function with complex argument~\cite[p.~447]{abramowitz-stegun}:
\begin{equation}
\label{gensol}
\begin{array}{rcl}
\mathcal{A}_h(z)&=&
 \kappa_1 J_{\nu e^{3\pi i/4}}\left(\nu^2 e^{-z_\star/\nu^2}\right)
 \Gamma\left(1+e^{3\pi i/4}\nu\right) \\
\noalign{\smallskip}
 &+& \kappa_2 J_{-\nu e^{3\pi i/4}}\left(\nu^2 e^{-z_\star/\nu^2}\right)
 \Gamma\left(1-e^{3\pi i/4}\nu\right).
\end{array}
\end{equation}
Here we have abbreviated $\nu=\sqrt{\omega/q c}$ and $z_\star=\omega z/c$.
The parameter $\nu$ is usually very large for typical frequencies and gravitational accelerations.
The electric field of the wave is then given by
\begin{equation}
\label{EfieldA}
\mathbf{E}(z,t)=\mathbf{E}_0\left(\mathcal{A}_h(z)+\mathcal{A}_p(z)+z\right) e^{-i\omega t}
=\mathbf{E}_0\mathcal{A}_h(z)\,e^{-i\omega t},
\end{equation}
because the particular solution of the non-homogenous equation, $\mathcal{A}_p(z)$ cancels out. 
\begin{figure}
\include{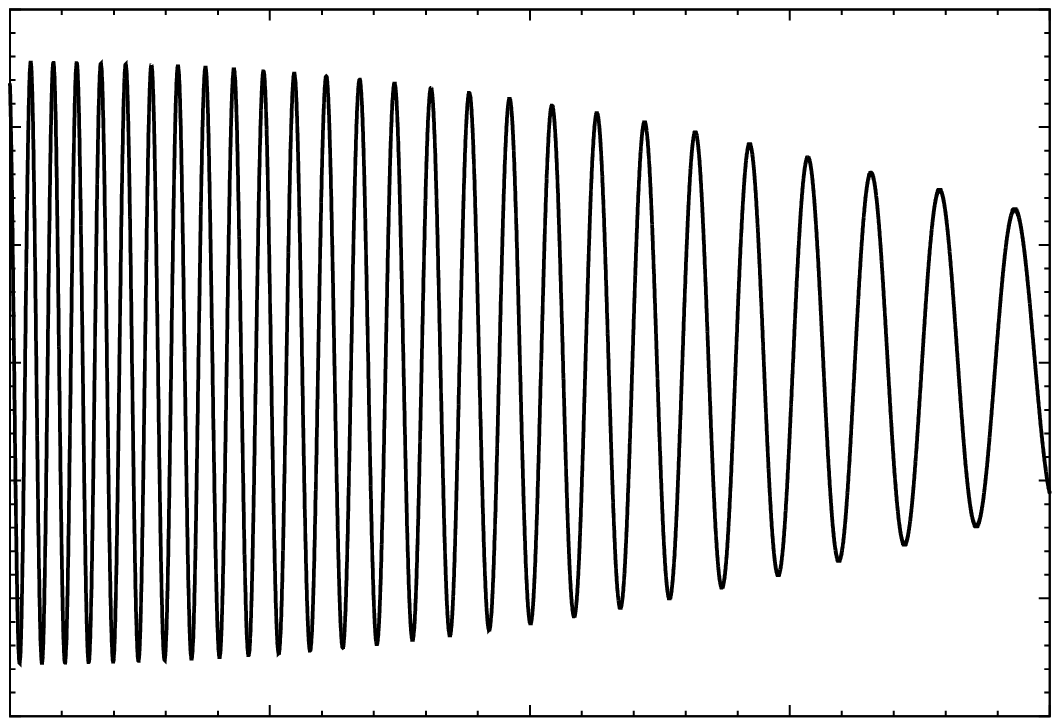}
\caption{Electric field amplitude as a function of scaled height, $z_\star$, for an electromagnetic
wave in a strong uniform gravitational field with $\nu=15$. The coefficients of $\mathcal{A}_h(z)$
are $\kappa_1=10^{13}$ and $\kappa_2=0$.\label{fig2}}
\end{figure}
In Fig.~\ref{fig2} we have plotted the electric field amplitude, which is obtained
from the real part of Eq.~(\ref{EfieldA}) after substituting Eq.~(\ref{gensol}). The numerical
parameters are chosen $\nu=15$, $\kappa_1=10^{13}$, and $\kappa_2=0$.

The damping of the wave is associated with the gravitational redshift of photons as discussed below.
Notice that the wavelength is also increasing and, reciprocally, the frequency is decreasing as
the wave travels upwards through the uniform gravitational field. The coefficient $\kappa_2$ must
be zero, because the second term in Eq.~(\ref{gensol}) corresponds to an amplification of the wave
as it travels upwards and, consequently, is unphysical.

\begin{figure}
\include{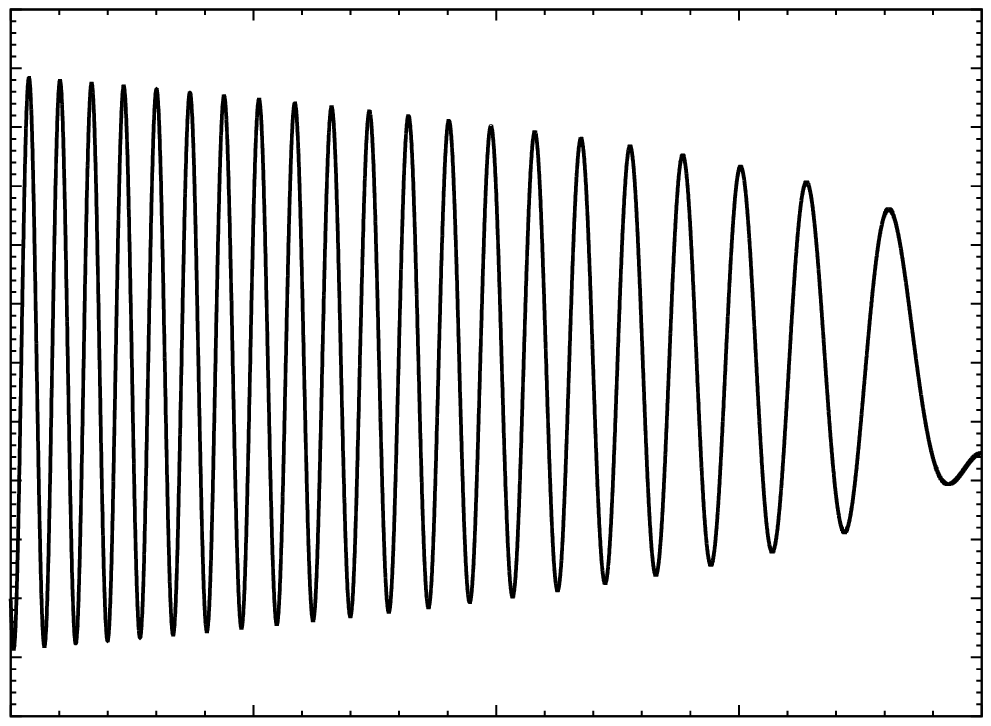}
\caption{Electric field amplitude as a function of scaled height, $z_\star$, for an
electromagnetic wave in a strong uniform gravitational field with $\nu=20$. We have
used the approximation in Eq.~(\protect\ref{Airy}) with $\kappa_1=1$ and $\kappa_2=0$.
\label{fig3}}
\end{figure}

On the other hand, as discussed in Sec.~\ref{Sect_I}, gravitational redshifts could be successfully
detected even for weak gravitational fields, such as the local field at the surface of the earth.
This justifies to also study the approximation of the general solution Eq.~(\ref{gensol}) for
the case of weak fields, with the limits $g \rightarrow 0$ or $\nu \rightarrow \infty$.
If we try to accomplish this task directly from Eq.~(\ref{gensol}), we will face some technical
difficulties because of the imaginary index of the Bessel functions. Moreover, the method of the
stationary phase is also difficult to apply because $\nu$ appears also as an argument of the
Bessel functions. It is far more convenient to start with the differential equation for
$\mathcal{A}(z)$, given by Eq.~(\ref{Aeq}), and carry out the expansion for $q=g/c^2\ll1$:
\begin{equation}\label{airyDE}
\displaystyle\frac{d^2 \mathcal{A}}{d z^2}
+ \left( \displaystyle\frac{\omega^2}{c^2}\big(1-2 q z\big)
+ i\displaystyle\frac{q \omega}{c}\right)\left(\mathcal{A}+z\right)=0.
\end{equation}
An explicit solution of the homogeneous equation, Eq.~\eqref{airyDE}, is now found in terms
of the Airy functions
\begin{equation}
\label{Airy}
\mathcal{A}_h(z) =
\kappa_1 \mathop{\rm Ai}\left(
\displaystyle{2 z_\star-\nu^2-i\over(2\nu)^{2/3}}
\right)
+ \kappa_2 \mathop{\rm Bi}\left(
\displaystyle{2 z_\star-\nu^2-i\over(2\nu)^{2/3}}
\right),
\end{equation}
where $z_\star=\omega z/c$ and $\nu=\sqrt{\omega c/g}$, as before. Again, out of physical grounds,
one has to take $\kappa_2=0$. Moreover, it can then be seen that only
the imaginary part of the Airy function of the first kind in Eq.~(\ref{Airy}) is physically
meaningful, because it corresponds to the damping of the wave travelling along the $z$-axis in
positive direction. The result is shown in Fig.~\ref{fig3} for $\nu=20$.

A further simplification may be obtained for $\nu \rightarrow \infty$ by using the following
asymptotic expansion of the Airy function for large $\vert \zeta \vert$ with $\zeta\in\mathbb C$, see
Ref.~\cite[p.~448, eq.~10.4.59]{abramowitz-stegun}:
\begin{equation}
\label{Aiapprox}
\mbox{Ai}(\zeta)=\displaystyle\frac{1}{2\sqrt{\pi}}\,\zeta^{-1/4} e^{-\frac{2}{3} \zeta^{3/2}}
\displaystyle\sum_{k=0}^\infty (-1)^k c_k \left(\displaystyle\frac{2}{3}\zeta^{3/2}\right)^{-k},
\qquad (|\arg\zeta|<\pi)
\end{equation}
where $c_0=1$ and $c_k=\Gamma(3 k+1/2)/\big(54^k k!\,\Gamma(k+1/2)\big)$.

As we are interested in recovering the classical result for
gravitational redshifts in a weak gravitational field, we can
safely ignore the algebraic prefactors in Eq.~\eqref{Aiapprox}.
The reason for this simplification is as follows: the vertical distance
between the emission point of the photon and the receiver in a
Pound-Rebka type experiment is much smaller than the characteristic length
associated with the approximately uniform field, $L=c^2/g\gg z$.
This implies that $z_\star\ll\nu^2$, and consequently
$$
\zeta = \frac{2z_\star-\nu^2-i}{(2\nu)^{2/3}}
= \left(\frac{2z_\star}{\nu^2}-1-\frac{i}{\nu^2}\right)\,\frac{\nu^{4/3}}{2^{2/3}}
= \frac{\nu^{4/3}}{2^{2/3}}\left(-1-\frac{i}{\nu^2}+O(z_\star/\nu^2)\right).
$$
The terms $O(z_\star/\nu^2)$ can then be regarded as the prefactors
multiplying the exponential term in Eq.~\eqref{Aiapprox}.

Ignoring all $\nu$-dependent factors which later on can be absorbed into the definition of
$\mathbf{E}_0$ and taking both possible roots in the exponential, Eq.~\eqref{airyDE} reduces to
\begin{equation}
\label{Aexp}
\mathcal{A}_h(z) \sim
e^{\pm\frac{i}{3} \nu^2
\left[1+(i-2z_\star)/\nu^2\right]^{3/2}}
\approx
e^{\pm\frac{i}{3} \nu^2
\left[1+\frac{3}{2}(i-2z_\star)/\nu^2+\frac{3}{8}(i-2z_\star)^2/\nu^4+O(\nu^{-6})
\right]^{3/2}},
\end{equation}
where we have expanded the exponent as a series of powers of $\nu^{-2}$ up to second order.
Finally, after choosing the negative sign and substituting $z_\star/\nu^2=gz/c^2$,
Eq.~(\ref{EfieldA}) gives the following result for a damped wave
\begin{equation}
\label{Edamped}
\mathbf{E}=\mathbf{E}_0 e^{-gz/2c^2}e^{i\omega (z/c-t)}.
\end{equation}
Notice that $t$ is the coordinate time as introduced in Eq.~\eqref{proposal}.
Here we have again included all remaining terms containing $\nu$ (but not in combination
with $z_\star/\nu^2$) in the amplitude $\mathbf{E}_0$.
A similar expression may be derived for the magnetic field. This solution represents
a damped electromagnetic wave in a conducting medium with conductivity $\sigma=g/(c^2\mu_0)$,
as could have been anticipated by inspection of Eq.~(\ref{Etelegraph}).
Due to the factor $e^{-2 g z/c^2}$ in the second-order time derivative in Eq.~\eqref{Etelegraph},
the second-order approximation contains a variable frequency~$\omega$.

Knowing the explicit form of the electric and magnetic fields put us in the position to be
able to calculate the energy density of the wave as a temporal average
over the coordinate time:
\begin{equation}
\label{energy}
\rho(z) = {1\over2}
\left\langle\varepsilon_0\mathbf{E}^2+\displaystyle\frac{\mathbf{B}^2}{\mu_0}\right\rangle
= {1\over2}
\left(\varepsilon_0 \mathbf{E}_0^2+\displaystyle\frac{\mathbf{B}_0^2}{\mu_0}\right)e^{-g z/c^2}
\left\langle \cos^2\left( \omega (z/c-t)\right) \right\rangle=\rho_0 e^{-g z/c^2}.
\end{equation}
Obviously the choice of time variable (coordinate or proper time) for the temporal average
in Eq.~\eqref{energy} can not influence the result for the energy density. If $T$ denotes
the wave period by a static clock in the coordinate frame, then the proper wave period in
the UGF frame is $\mathcal{T}=\sqrt{-g_{00}}\,T$. It is then easy to see that 
the average in both frames for one wave period $T$ or $\mathcal{T}$, respectively, yields
the same factor $1/2$.

Note that the same exponential factor also appears in the expression for the frequency due
to standard gravitational redshift
\begin{equation}
\label{gravred}
\omega(z)=\frac{\omega_0}{\sqrt{-g_{00}}}=\omega_0 e^{-g z/c^2},
\end{equation}
where we have used the metric Eq.~\eqref{metric} in the UGF system defined in
Eq.~\eqref{alpha}.

Therefore, Eqs.~\eqref{energy} and \eqref{gravred} demonstrate that the ratio of energy
density and frequency of the wave travelling through the uniform gravitational field is
always constant, regardless of its height $z$ as measured by a static receiver at this position.
If $n$ is the average number of photons
per unit volume, their corresponding energy density is $\rho=n\hbar\omega$ according to
Planck's fundamental postulate of quantum mechanics. It is then clear that
\begin{equation}
\label{Planck}
\displaystyle\frac{\rho}{\omega}=\displaystyle\frac{\rho_0}{\omega_0}= n\hbar.
\end{equation}
Hence, we observe that general relativity is compatible with Planck's postulate concerning
the interpretation of the redshift in a strong uniform gravitational field from the point of
view of the covariant Maxwell equations in a curved spacetime.

\section{Conclusions and Remarks}
\label{Sect_IV}

In this paper we have studied the solutions of Maxwell equations in a uniform
gravitational background field or, alternatively, in a uniform accelerating rigid frame. We have
shown that explicit solutions can be found for electrostatic fields produced by an evenly
charged metallic plate and for the case of electromagnetic waves in the vacuum. The wave
equation in a gravitational field is analogous to the telegraph equation obtained in classical
electrodynamics when electromagnetic waves propagate in a conducting media. However, the
conductivity $\sigma=g/(\mu_0c^2)$ of such a medium, caused for example by the gravitational
field of a typical celestial body, is extremely small. For any feasible test frequency $\omega$
of light, this corresponds to a very large penetration depth of the associated electromagnetic
waves.

Moreover, we must recognise that uniform gravitational fields are an idealised case and as such
not found in nature. The Schwarzschild metric would be the adequate framework to study realistic
gravitational fields. Nevertheless, even as a local approximation for the field near the surface
of the earth, the solutions found provide a useful insight into the behaviour of electromagnetic
waves and photons in a gravitational field. In particular, we could show that the amplitude is
described in terms of the Airy function depending on height.

The ratio among the energy density of the electromagnetic wave and its frequency is fixed to
a constant as the wave travels across the gravitational field. This constant is proportional
to Planck's constant, which is to be expected by compatibility with Planck's fundamental
postulate. Any other result would pose a serious contradiction between the general theory
of relativity and quantum mechanics. A similar coherence between both theories is also found
for the Compton and Doppler effect in special relativity as studied from a kinematic point of
view (taking into account the recoil of a massive body which emits photons)~\cite[pp.~194]{french}.

The relation among quantum phenomena and gravitation in earth-bound experiments goes beyond
mere academic or pedagogical interest. For example, the recent proposal for a probabilistic
description of gravity, the so-called entropic theory of gravity~\cite{verlinde}, has been
argued to fail for the description of the aforementioned experimental results for
quantum states of ultracold neutrons in the earth's gravitational field~\cite{kobakhidze}.
For these reasons, it would be interesting to analyse the behaviour of experimentally viable
quantum states with background Schwarzschild or Kerr metrics as a way to unveil or predict
some further, hopefully surprising, connections between classical gravity and the microscopic
world. Apart from employing Maxwell's covariant equations, it would also be interesting to
analyse predictions of quantum field theory on a fixed background. The WKB approximation
would be applied to obtain results in a general metric. Work along these lines is in progress.


\end{document}